\documentclass[acmsmall, fleqn, nonacm]{acmart}

\AtBeginDocument{}

\setcopyright{acmlicensed}
\copyrightyear{2025}
\acmYear{2025}
\acmDOI{XXXXXXX.XXXXXXX}

\usepackage{hyperref}
\usepackage{stmaryrd} 
\usepackage{tikz}
\usepackage{xcolor}

\newcommand{\Bool}{\mathbb{B}}
\newcommand{\Nat}{\mathbb{N}}
\newcommand{\Pos}{\mathbb{N}^+}
\newcommand{\Int}{\mathbb{Z}}
\newcommand{\Real}{\mathbb{R}}

\font \aap cmmi10 
\newcommand{\at}[1]{\mbox{\aap ,} #1}
\newcommand{\true}{\textit{true}}
\newcommand{\false}{\textit{false}}

\usepackage{array}
\newcolumntype{L}{>{$}l<{$}}
\newcolumntype{C}{>{$}c<{$}}
\newcolumntype{R}{>{$}r<{$}}

\newenvironment{mcrl2}{%
  \begin{trivlist} \item \mbox{} \qquad \qquad
    \begin{tabular}{@{}>{\bf}p{2.3em}L@{\ }L@{\ }L@{\ }L@{\ }L@{\ }L@{\ }L}
}{%
    \end{tabular}
  \end{trivlist}
}

\newcommand{\added}[1]{{{\color{blue}{#1}}}}
\renewcommand{\added}[1]{#1}

\begin{document}

\title{A Comprehensive History of $\mu$CRL and mCRL2}

\author{Jan Friso Groote}
\email{j.f.groote@tue.nl}
\orcid{0000-0003-2196-6587}

\author{Erik P. de Vink}
\email{e.p.d.vink@tue.nl}
\orcid{0000-0001-9514-2260}

\affiliation{%
  \institution{Eindhoven University of Technology}
  \country{The Netherlands}
}

\renewcommand{\shortauthors}{Groote and De Vink}

\begin{abstract}
  \noindent
  This article gives a historical overview of the background,
  motivation and development of $\mu$CRL and its successor mCRL2, from
  the inception to the present.
  Both mCRL2 and $\mu$CRL are similar, compact, but very expressive
  formalisms based on process algebra, term rewriting, and the modal
  mu-calculus. They are developed to model and analyse the behaviour
  of interacting systems, i.e., \added{systems that communicate by
    exchange of messages, among each other and with the outside
    world}.
  Every contemporary computer system can be viewed as such an
  interacting system and their communication schemes are difficult to
  design correctly.
  By sticking to the mathematical foundations, but being led by the
  desire to be practically relevant, the formalism has grown to become
  very versatile.
  In particular, mCRL2 does not only foster the development of theory
  and the formulation of correctness proofs, but it is also the basis
  of a large set of automatic tools that help to provide insight in
  the behaviour of complex computer controlled systems.
\end{abstract}

\ccsdesc[500]{Theory of computation~Logic and verification}
\ccsdesc[500]{Theory of computation~Modal and temporal logics}
\ccsdesc[300]{Theory of computation~Equational logic and rewriting}
\ccsdesc[500]{Theory of computation~Verification by model checking}

\keywords{Formal Methods, History, $\mu$CRL, mCRL2, Process Algebra,
  Abstract Data Types, System Analysis and Verification, Model Checking}

\settopmatter{printacmref=false}
\setcopyright{none}
\renewcommand\footnotetextcopyrightpermission[1]{}
\pagestyle{plain}

\maketitle

\section{Introduction}

The mCRL2 modelling language and toolset, together with its
predecessor $\mu$CRL, have been in existence for more than three
decades now. They are based on the basic concept of an atomic action
in the style of Milner~\cite{DBLP:books/sp/Milner80}, use abstract
data types, and allow to perform behavioural analysis by modal
logics. Together this constitutes a very stable, mathematically
well-founded, expressive, and versatile environment to describe and
analyse the behaviour of communicating systems.

Specification formalisms most similar to mCRL2 and $\mu$CRL are
CADP~\cite{CADP96:cav, GLMS13:sttt} and FDR~\cite{Ros94:hoare,
  Ros97:ph}. The former puts forward a more programming style of
specification, whereas the latter lacks the support of modal
logic. Besides these, there are many other specification formalisms,
but they all are less expressive and generally concentrate on
particular verification techniques or particular specification
aspects.

The advent of mCRL2 and $\mu$CRL cannot be seen as a well planned
act. There were three ingredients that came together. The first one
was the existence of the research group AP2, a subgroup of the
Afdeling Programmatuur, at the Centre of Mathematics and Computer
Science (CWI, Centrum voor Wiskunde en Informatica), together with
researchers at the University of Amsterdam and the Vrije Universiteit,
all in Amsterdam.  These groups had grown substantially through
European subsidies.

The groups were strongly influenced by Bergstra. His interest in
the process algebra ACP as well as in abstract datatypes was the
second determining factor, and these were the topics the group members
worked on. Most of the group members had a background in mathematics,
and there was an absolute desire to apply mathematics at the highest
standards.  This did not only imply that mathematics had to be applied
with appropriate rigour, but it also meant that the developed theories
should be as simple and elegant as possible.  Related research groups
at CWI were AP1, led by De~Bakker, which was focussed on metric
semantics of programming languages~\cite{BV96:mit}, and AP3, led by
Klint, which primarily focussed on language workbenches, strongly
inspired by equational abstract data types and rewriting
\cite{BHK89:acm}.

The third factor was the desire to show that the developed theory was
useful for practical software engineering. Initially, process algebra
in Amsterdam came into being to find a simple way to understand the
semantics of recursive programs, answering a question by De~Bakker.
Abstract data types were motivated by the desire to become independent
of mathematically awkward machine dependent data types.  Later came
the desire to show that especially process algebra could be used to
understand and prove the correctness of distributed systems and
protocols.

It was uniformly observed that process algebra alone was insufficient
to describe all aspects of behaviour, and it came natural to extend it
with equational abstract data types. This was done in
LOTOS~\cite{LOTOS}, PSF \cite{MV90:fi,mauwPSF}, and later in the
Common Representation Language (CRL), developed within the RACE SPECS
project. Being unsatisfied by the complexity of CRL, a more concise
variant was developed called $\mu$CRL, being a quite minimal
combination of process algebra with equational datatypes. Care was
taken at the outset that $\mu$CRL was mathematically elegant and had a
well-defined syntax and semantics.

Around 2000 activities began to define mCRL2 to remedy a few
shortcomings in $\mu$CRL.  This work took place at Eindhoven
University of Technology, while in the mean time at CWI and Vrije
Universiteit \added{Amsterdam} $\mu$CRL continued to be supported for
some time.  In Eindhoven, the modal mu-calculus was added to mCRL2 to
formulate properties.

\added{Initially, the practical use of $\mu$CRL and mCRL2 was its
  proof system; correctness proofs showing the equivalence of a
  compact specification and a realistic implementation were
  constructed by hand.}
\added{In the behavioral setting, processes are considered
  equivalent if no difference can be observed between them. As there
  many ways to observe processes, there are many notions of process
  equivalence~\cite{Gla90:concur, Gla93:concur}. Within $\mu$CRL and
  mCRL2 the most frequently used process equivalences are strong,
  weak, and branching bisimilarity \cite{DBLP:conf/tcs/Park81,
    DBLP:books/sp/Milner80, DBLP:journals/jacm/GlabbeekW96}.}
Some of \added{the obtained} correctness results have been
verified by proof checkers such as Coq
\cite{DBLP:journals/fac/BezemBG97, FGPBP04:amast,
  DBLP:journals/fac/GrooteMS05}. Later attention focussed on the use
of automated analysis of correctness using the toolset, as this is far
more efficient and applicable too much larger models.

We believe that we have been successful in defining useful languages,
first in the form of $\mu$CRL and later in the shape of mCRL2.
This is the reason why mCRL2 is  ---after such a long time--- still
quite alive and kicking, being applied to model and analyse a wide
range of systems, and being used as the basis to develop more and also
more advanced behavioural analysis techniques.

This document describes the history of both $\mu$CRL and~mCRL2 in
quite some detail.  We focus on the motivations and do not provide a
technical account. For this for instance
\cite{DBLP:books/mit/GrooteM2014, DBLP:series/txtcs/Fokkink07} are
more appropriate.  Although mCRL2 and its surrounding toolset have
become a very useful framework for the analysis of computer
interactions, there are still numerous unanswered questions that lead
to further investigation. For instance, an important question is how
to analyse even more complex systems, especially, when containing
probabilities, time, or even interaction with the continuous physical
world. \added{All of this takes place in a world where
  computer-controlled systems have grown more complex over time,
  handling a wide variety of related and unrelated tasks
  simultaneously and relying on processors that execute many
  operations in parallel in line with relaxed memory models.}  Keeping
all this complexity under control remains a daunting task, in which,
as we believe, formalisms like mCRL2 can play important roles.

\section{A Common Representation Language}

In this section we describe how the Common Representation Language
came into being.  While being developed this language was already
considered to be unmanageably complex, leading to a micro variant of
it, called $\mu$CRL. The basic ingredients of CRL, $\mu$CRL, and also
mCRL2 are process algebra and abstract data types. From a historical
perspective, this was a natural combination, in line with natural
developments of computer science in the eighties.

The end of the second world war saw a burst of development in
automated computers.  These had to be programmed in assembly code
which was a delicate activity.  As a solution higher-level languages
were developed like Algol~60, Cobol, and Fortran.  Contrary to
expectation \cite{Alberts2023}, higher-level programming languages did
not turn out to be the solution for what was later called the software
crisis \cite{Bro75:aw}.  Solutions were searched in developing better
notions to use in such programming languages, precisely defining the
semantics of such languages, and developing formalisms using which
programs could be described more abstractly.

In 1973, Robin Milner published two papers outlining the need to base
the semantics of programs on actions~\cite{Mil73,Mil73it}. In
addition, he defended the use of parallelism and non-determinism.
These observations led to the Calculus of Communicating Systems (CCS),
a basic process modelling language supported by bisimulation as the
notion of process equivalence enjoying an
axiomatisation~\cite{DBLP:books/sp/Milner80}.

The work of Milner strongly influenced the development of the language
of Communicating Sequential Processes (CSP) by Hoare, who in 1978
presented~CSP as a Dijkstra-style parallel programming
language~\cite{DBLP:journals/cacm/Hoare78}, but in 1985 CSP~had become
a process calculus quite in the style of
Milner~\cite{DBLP:books/ph/Hoare85}.
Klop and Bergstra were working in Amsterdam and sought
to understand the semantics of recursive programs. They came up with
the Algebra of Communicating Processes (ACP), which is very much alike
CCS, although they claimed that they only later learned about
it~\cite{DBLP:journals/iandc/BergstraK84}. In the introduction
of~\cite{Groote91} a more extensive history of the early days of
process algebra is given, also as an overview of some other process
formalisms that existed at the time.

The early process calculi CCS and~ACP, though simple and elegant by
nature, were lacking data types and, consequently, were rather limited
in expressivity.
Aiming to leverage manual verification of small-scale programs to
validation of industrial-size software, process calculi were to be
combined with abstract data types.
The development of a formal theory of abstract data types goes back to
the late seventies, see~\cite{Zil74:mit, GTW78:current, EKP79:sigact}
for example.
Among others, Goguen and Meseguer advocated in their work on~OBJ a
computational interpretation of initial algebra semantics by viewing
the equational specification of data as a term rewriting
system~\cite{Gog87:ctrs}.
By putting restrictions such as confluence and termination on the
rewrite systems, normal forms uniquely identify classes of equivalent
terms.

The idea of such an integration with abstract data types was first
undertaken in the language LOTOS~\cite{LOTOS}, which became an
international standard. LOTOS was intended as an abstract
specification formalism, not as a programming language.
LOTOS builds on Milner's CCS and Ehrig's algebraic specification
language ACT ONE~\cite{EM85:fund}.
For~ACP, Mauw and Veltink developed the specification formalism PSF,
extending the process calculus~ACP with means to incorporate abstract
data types along the lines of the algebraic specification language
ASF~\cite{BHK89:as} that was based on work by Bergstra and Tucker.
The languages LOTOS and~PSF are very similar in nature to $\mu$CRL and
mCRL2, but less concise as they provide a broader variety of language
constructs and have a verbose syntax.
LOTOS has been taken as the basis for LOTOS-NT, or~LNT, a transition
nicely described in~\cite{DBLP:conf/birthday/GaravelLS17}.

\begin{figure}[ht]
\begin{center}
\begin{tikzpicture}[shorten >=2pt, >=stealth]

\draw [thick, <-] (1,1) -- +(0,0.5) node[above] {LOTOS};
\draw [thick, <-] (3,1) -- +(0,0.5) node[above] {Chill};
\draw [thick, <-] (5,1) -- +(0,0.5) node[above] {PSF};
\draw [thick, <-] (7,1) -- +(0,0.5) node[above] {SDL};
\draw [thick] (0,0) -- +(8,0) -- +(8,1) -- +(0,1) -- cycle +(4,0.5) node {CRL: Common Representation Language};

\draw [thick, ->] (1,0) -- +(0,-0.5) node[below] {Typechecker};
\draw [thick, ->] (3,0) -- +(0,-0.5) node[below] {Simulator};
\draw [thick, ->] (5,0) -- +(0,-0.5) node[below] {Visualiser};
\draw [thick, ->] (7,0) -- +(0,-0.5) node[below] {\hspace*{1cm}State space generation};

\end{tikzpicture}
\end{center}

\caption{Language translations and tools in the SPECS project}
\label{fig1}
\Description{A central box representing CRL with incoming arrows from
  above for every specification language and outgoing arrows to below
  for every desired tool}
\end{figure}
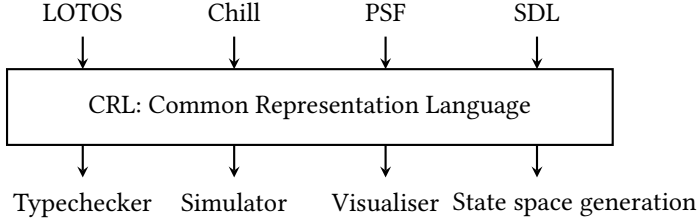

From 1989 until 1992, the European project SPECS (in full, RACE
contract 1046, Specification and Programming Environment for
Communication Software) took place. Its central idea was that many
specification languages had emerged, e.g., LOTOS \cite{LOTOS}, PSF
\cite{mauwPSF}, CHILL \cite{Chill}, SDL \cite{SDL}, and all these
languages required further support by various kinds of tools like
simulators, visualisers, state space generators, etc. So, with
$n$~languages and $m$~kinds of tools, the momentous effort of building
$n\times m$~pieces of software had to be undertaken. But, by defining
a central \textit{Common Representation Language} (CRL), to
which every language could be translated and for which only a single
instance of each kind of tool needed to be made, \added{the work}
would substantially reduce to \added{the construction of $n+m$
  artefacts,} viz.\ $n$~translators to CRL and $m$~tools for CRL
itself. The idea is depicted in Figure \ref{fig1}.  

The definition of CRL was initially taken up rather
seriously~\cite{CRL90}. Both its syntax and complete semantics were
carefully defined. However, as CRL had to accommodate all semantical
constructs of the various input specification languages, it quickly
became so complex that its definition became impossible to maintain.
This also meant that building tools that would cover CRL completely
would be an impossible task.
\added{Even worse, as obvious constructs in a source language became
  obfuscated by the translation to CRL, building tools for one single
  source language in this framework was also harder. }
Actually, no translator nor tool for CRL
was ever finished.

The conclusion at this point was that because of its complexity a
language like CRL could never become successful. The only viable route
forward would be to define a very concise language, with a minimum of
language constructs, which at the same time would be highly
expressive. The proposed language was baptised micro-CRL or
$\mu$CRL\@. It essentially consisted of the process algebra ACP to
express behaviour, extended with equationally defined abstract data
types for the data. Its complete syntax and semantics are described in
a compact report \cite{GP91,Groote91,GP90}. Although we were convinced
that this was a wise, and actually the only reasonable
step towards a useful common representation language, the change to
$\mu$CRL was not appreciated within the SPECS project. Outside of
SPECS, $\mu$CRL started a life of its own.

\section{$\mu$CRL}

As indicated, the language $\mu$CRL is a compact language that allows
to describe interacting processes with data. A textbook describing the
language is~\cite{DBLP:series/txtcs/Fokkink07}. Deviating from the
original motivation of CRL, $\mu$CRL became a basic, but
mathematically precise modelling language to characterise and study
the behaviour of interacting systems. Its initial use was to manually
prove the correctness of \added{distributed} algorithms \added{
  and communication protocols}, for which a dedicated proof system was
developed~\cite{GP93:sosl}. Later came the desire to automate the
analysis. In this section we restrict ourselves to the language,
leaving the analysis part to later sections.

\subsection{Behavioural specification in $\mu$CRL}

The language $\mu$CRL is built up around actions that have a short or
indicative name like $a$ or $\textit{send\_message}$. Actions can have
parameters, like $\textit{send\_message} \mkern1mu (3,[ \mkern2mu ])$
where $[ \mkern2mu ]$ represents the empty list. Actions represent
atomic events, like sending the empty list to recipient~$3$, but can
also stand for receiving a message, instructing a robot, or even
something physical like starting to open a door. If an action~$a$
takes place, it will terminate afterwards. Actions can also be used
for communication: if two components perform two actions, say one is
receiving and the other sending, then they can happen synchronously
and this represents communication. There is one special action, the
hidden action~$\tau$, which can take place but cannot be directly
observed.

In the process part of $\mu$CRL, actions are combined using the
sequential operator~(`$\cdot$'), the choice operator~(`$+$'), the
parallel operator~(`$\parallel$'), the conditional operator
(`$\_\triangleleft\_\triangleright\_$') and the constant~$\delta$. A
sequential composition~$p{\cdot}q$ behaves as the process~$p$ and if
$p$~terminates allows the behaviour of~$q$ to happen subsequently. A
choice~$p+q$ indicates that the behaviour of~$p$ or the behaviour
of~$q$ can be done where the first action that happens in $p$ or~$q$
determines which of the two is chosen. So, the process
$a{\cdot}b+c{\cdot}d$ indicates that it is possible to do either
an~$a$ followed by a~$b$, or a~$c$ followed by a~$d$. The parallel
composition~$p \parallel q$ indicates that the behaviours of $p$
and~$q$ are executed in parallel, where the atomic actions in~$p$
and~$q$ \added{take primarily place in an interleaved fashion, while
  certain actions in $p$ and~$q$ can happen synchronously}. There are
also the left merge operator (`$\llfloor$') and the communication
merge operator (`$\mid$') which have a technical purpose, which we
skip here.  The conditional $p \triangleleft c \triangleright q$ is
the then-if-else construct in CSP-style notation. If condition~$c$ is
true, process~$p$ is executed, and if~$c$ is false, then process~$q$
takes place. The process~$\delta$ is a process that cannot do
anything, not even terminate, and is therefore called deadlock or
inaction.
\added{In~CSS, $\delta$~was initially denoted as NIL and later
  as~$0$~\cite{DBLP:books/sp/Milner80, Mil90:ph}, and in CSP and LOTOS
  it was written as
  STOP~\cite{DBLP:series/txcs/Roscoe10,LOTOS}. Notably, $\delta$~does
  occur in the semantics of LOTOS, but there it is used as a means to
  return results of processes.}

The choice operator can be used to model inputs. For example, the
process
$\textit{read} \mkern1mu (\textit{true}) {\cdot} \mkern1mu p +
\textit{read} \mkern1mu (\textit{false}) {\cdot} \mkern1mu q$ models
that a boolean is read. If the boolean is true, behaviour~$p$ is
executed, otherwise, behaviour~$q$ takes place. In order to also
express reading data from large or unbounded domains as a
generalisation of the choice operator, the sum operator
$\sum_{d:D} \,p$ is provided. Here, $d$~is a variable ranging over
domain~$D$. The variable~$d$ can occur in the process~$p$. It is
expressed that $p$~can take place for any value for~$d$. For instance
reading a number~$n$ to be used in process~$P(n)$ can be written by
$\textstyle\sum_{n:\Nat}\, \textit{read} \mkern1mu (n){\cdot}P(n)$.

\added{In process algebras, parallel processes communicate by
  letting actions synchronise, i.e., by letting them happen at the
  same time.  In~CCS an action~$a$ can synchronise with its
  co-action~$\overline{a}$ and the result of this synchronisation
  is the internal action $\tau$. By applying an explicit restriction
  operator, actions are forbidden to take place on their own, being
  forced into communication.
  In~CSP a similar scheme was used where actions were always forced to
  synchronise with actions with the same name if this action would
  occur in both parallel processes.

  For~mCRL2 a more flexible scheme was chosen, based on the
  communication function~$\gamma$ in~{ACP}. Any two actions are
  allowed to synchronise to an arbitrary action by declaring lines of
  the form
  \begin{displaymath}
    \textbf{comm}~a\mid b = c
  \end{displaymath}
  This says that action~$a$ can synchronise with action~$b$ and the
  result is action~$c$.  If actions have data, the communication is
  declared on the action labels, and synchronisation can only take
  place if the parameters of the actions are equal. The declarations
  are taken to be commutative, in the sense that with the line above,
  action~$b$ also synchronises with action $a$ to action~$c$. This
  implies that the parallel operator~$\parallel$ is commutative
  which is a required property of this operator.

  The communication declarations only indicate that actions are
  allowed to synchronise. In order to enforce synchronisation, the
  encapsulation operator $\partial_B(p)$, with~$B$ a set of action
  labels, is used to express that the actions in~$B$ are blocked from
  happening in~$p$. If $a$ and~$b$ are declared to synchronise to~$c$,
  encapsulating $a$ and~$b$ enforces these actions to synchronise
  to~$c$. Similar to the encapsulation operator, there is a hiding
  operator that renames actions to~$\tau$, and a general renaming
  operator, that can rename action labels to other action labels.

  The communication scheme in $\mu$CRL was problematic, and it was
  therefore replaced in mCRL2. The reason is as follows. The parallel
  operator is not only commutative, but also associative. This meant
  that communication declarations such as
  \begin{displaymath}
    \textbf{comm}~a \mid b = c,\  c \mid d = e
  \end{displaymath}
  were not allowed as they would lead to a non-associative parallel
  operator. This was checked by the tools.  It was necessary to extend
  this communication declaration to
  \begin{displaymath}
    \textbf{comm}~a \mid b = c,\ c \mid d = e,\ b \mid d = f,\ a \mid f = e
  \end{displaymath}
  which included the introduction of a dummy action $f$. For
  synchronising more than three actions, such declarations became
  forbiddingly large, effectively rendering $\mu$CRL unusable to
  specify multi-party synchronisation.
} 

Recursive behaviour is expressed in $\mu$CRL by guarded process
equations, e.g., $X=a{\cdot}X$. This process equation expresses that
process variable~$X$ can perform the action~$a$ after which it behaves
as~$X$ again. This means that $X$~represents the process that can do
action~$a$ indefinitely. The equation is guarded since the action~$a$
precedes the process variable. The guardedness guarantees unique
solutions of such equations~\cite{Ard61:swct}. The process variables
can be parameterised with any arbitrary number of parameters. For
example, the equation
$X(n \mkern1mu {:} \mkern1mu \Nat) = a(n){\cdot}X(n \mkern1mu {+}1)$
has a unique solution for~$X$ such that $X(0)$~can do the actions
$a(0) \mkern1mu a(1) \mkern1mu a(2) \mkern1mu a(3) \ldots {}$,
consecutively.

The behavioural part of $\mu$CRL was later extended with
time~\cite{gro97}. The at operator~$\at{}$ as used in~$p\at{t}$
expresses that the first action in~$p$ has to take place at
time~$t$. Following the approach for timed
I/O~automata~\cite{LT89:cwi, KLSV06:mc} and that of
Uppaal~\cite{LPY97:sttt, LLN18:fm}, the semantics allows consecutive
actions at the same moment in time. E.g., the process
$a\at{1} {\cdot} \mkern1mu b \mkern1mu \at{1}$ can execute the
action~$a$ at time~1, followed by the action~$b$ also at
time~1\@. Executing actions backward in time is not allowed. For
instance, $a\at{2} {\cdot} \mkern1mu b \mkern1mu \at{1}$ is equal to
$a\at{2} {\cdot} \mkern1mu \delta\at{2}$ indicating that after the
action~$a$ at time~$2$ a time deadlock at time~$2$ occurs, meaning
that after the action~$a$ at~$2$ nothing can happen anymore, including
the progress of time.

\subsection{Data types in $\mu$CRL}

The data types in $\mu$CRL were initially very simple. Data domains
could be declared using a keyword $\textbf{sort}$. The keyword
$\textbf{func}$ was added to declare functions \added{using which data
  expressions could be constructed}. Using the keyword $\textbf{rew}$
equations were provided to describe \added{which terms should be
  considered equal}. More specifically, the keyword $\textbf{rew}$
indicates an implicitly intended direction to evaluate expressions
using term rewriting \added{\cite{Terese03}}, although the semantics
of the data types treats the equations as pure non-directional
equations.

Each data type \added{in $\mu$CRL} must be defined explicitly, where
each specification contains at least the sort $\textbf{Bool}$
containing the two constructors $\textit{T}$ and $\textit{F}$,
representing true and false.  Also, all operators on the domains, like
\textit{and}, \textit{or}, etc., have to be provided
explicitly.
\added{Typically, a $\mu$CRL specification would start with
\begin{mcrl2}
   func & T,F: \textbf{Bool} \\
        & \mathit{and}:\textbf{Bool} \times\textbf{Bool}
        \rightarrow\textbf{Bool} \\
   var & x:\textbf{Bool} \\
   rew & \mathit{and}(T,x)=x \\
       & \mathit{and}(F,x)=F \\ 
\end{mcrl2}
}
\sloppy{%
  It is not uncommon for $\mu$CRL specifications to consist largely of
  data type definitions, typically defining booleans, natural numbers,
  lists, and so on.  \added{All} function symbols were denoted using
  prefix notation, typically leading to less readable expressions.
  For example
  $\textit{and} \mkern1mu (\textit{larger} \mkern2mu
  (x,\textit{zero}), \textit{larger} \mkern2mu (\textit{succ}
  \mkern1mu (\textit{succ} \mkern1mu (\textit{zero})),x))$.
} 

The model for the datatypes is an algebra of which the elements
correspond to the terms that can be made with the functions declared
with the keyword $\mathbf{func}$, that satisfies all
equations. \added{In addition, the two elements in the semantic
  domain of the booleans must be different, i.e.\ $T\not=F$. An
  algebra satisfying this requirement is called \textit{boolean
    preserving}}. The latter means that the domain of the booleans
\added{consists of exactly} two different elements represent true
and false. Note that this differs from initial
algebra models used in PSF and {LOTOS}. The semantics of the data types
is much more in line with loose semantics as proposed in
\cite{DBLP:journals/acta/GuttagH78}.

The form of a data type specification was subtly changed
in~\cite{gro97}, inspired by attempts to prove correctness of
distributed systems based on the process axioms and the
\added{definitions} of the abstract data types. \added{By
  declaring all functions in the same way, no distinction could be
  made between the constructors of a domain, such as $T$ (true) and
  $F$ (false), and those with a supporting role, such as
  $\mathit{and}$.}  Therefore, it was made possible to
\added{distinguish between} auxiliary mappings besides
constructors. So, instead of being forced to define the function
\textit{and} on booleans as an extra constructor of the domain, it
could now be declared as the mapping
\begin{displaymath}
  \textbf{map} \  \: \textit{and} :
  \textbf{Bool}\times\textbf{Bool}\rightarrow \textbf{Bool}.
\end{displaymath}
This mapping \textit{and} would not span up the data model. Only
constructors declared using the keyword $\textbf{func}$ would do so.
\added{ In this way it became explicit that proving properties on a
  data type could be done with induction over the constructors only.}
The semantics was changed to reflect this. In a model, all elements in
a domain for which constructors are defined must be denotable using
these constructors. If a domain has no constructors, there is no
constraint on the size of its model.

In this setting reasoning about data types is done using three
principles, namely, equational reasoning, deriving a contradiction by
showing that $\textit{T}=\textit{F}$, and induction on constructors.
\added{The} constructors of the booleans are
$\textit{T}$ and $\textit{F}$, and this leads to the induction
principle that if a formula is proven for $x=\textit{T}$ and
$x=\textit{F}$, then the formula holds for all booleans~$x$.
\added{This provided a quite adequate framework to specify most
  datatypes in a precise and concise manner. For instance, the natural
  numbers are neatly specified using Peano arithmetic, with
  constructors $0$ and $\mathit{successor}$, and lists used the empty
  list and the prepend operator as constructors. }

\section{mCRL2}

Around the year 2003 it was felt that $\mu$CRL did fall short in
several aspects \cite{DBLP:journals/entcs/GrooteMWU06}.
\begin{enumerate}
\item Having to specify all data types in each specification
  distracted from interesting behavioural phenomena. The use of
  functions in prefix notation, leading to expressions like
  $\textit{and}(b,\textit{T})$, did not help readability. The equational
  data types did not allow higher-order functions, although
  these can be quite helpful in specifications.
\item The way communication of actions was specified, using the
  $\textbf{comm}$ keyword, was not very appealing, especially because
  of its global scope over the whole specification and due to the fact
  that associativity of communication had to be specified explicitly,
  preventing a compact specification of multiparty synchronisation.
\item The then-if-else construct $p \triangleleft c \triangleright q$
  proved difficult to read; for comprehension of the behaviour of~$p$
  it helps if~$c$ is read first, but because $c$ follows~$p$,
  $c$~itself, especially when $p$~is large, may be hard to find.
\item The behaviour of timed actions that would occur simultaneously
  in sequence, e.g., $a \at{1} {\cdot} b \mkern1mu \at{1}$, 
  did not appear in line with the way time is working.
\end{enumerate}
Because addressing these issues would lead to notable alterations of
$\mu$CRL, it was decided to change the name to mCRL2. The Greek
letter~$\mu$ was changed into the letter~`m', as $\mu$~is
inconvenient for internet searches, which by then became
commonplace. The~2 was added to indicate that the language did change,
although the languages $\mu$CRL and mCRL2 remained not very far
apart. The book~\cite{DBLP:books/mit/GrooteM2014} describes mCRL2 in
extenso.

The various adaptations regarding data types were the most
substantial, although the data types in mCRL2 are in essence still
equational abstract data types, but now higher-order functions and
conditional rewrite rules became available.
The first change was the adaptation of keywords. Instead of
$\textbf{func}$ for constructors, the keyword $\textbf{cons}$ was
introduced to stress that there are now constructors and ordinary
mappings. 
Also, the keyword $\textbf{rew}$ was replaced by $\textbf{eqn}$ to
stress that the semantics of the data types considers the defining
equations as proper equations and does not make any reference to a
rewrite direction. Still, term rewriting is extensively used in the
mCRL2 tools to calculate that certain terms are equal, but from the
perspective of the semantics of the data types, this is only one way
to evaluate terms and any other technology respecting equations can be
used equally well. The other keywords such as $\textbf{sort}$ and 
$\textbf{map}$ did not alter.

A significant change was to allow function sorts. The sort
$D \rightarrow E$ represent the sort of all functions from sort~$D$ to
sort~$E$ where both $D$ and~$E$ can be function sorts themselves. This
also \added{requires} higher-order variables, i.e., variables that
range over functions. For example, the commonly known function
$\mathit{maplist}$ on lists can be defined in mCRL2 as follows.
\begin{mcrl2}
\textbf{map} & \mathit{maplist}:(D\rightarrow D) \times \textit{List}(D)
\rightarrow \textit{List}(D) \,;
    \\[3pt]
var & f:D\rightarrow D \,;\\
    & \ell:D \,;\\
    & L:\textit{List}(D) \,; 
      \\[3pt]            
\textbf{eqn} & \mathit{maplist}(f,[ \mkern2mu ])=[ \mkern2mu ] \,; \\
    & \mathit{maplist}(f,\ell \triangleright L)=f(\ell )\triangleright
      \mathit{maplist}(f,L) \,;
\end{mcrl2}
where $D$~is an arbitrary data type, $\textit{List}(D)$~is the type of
lists over~$D$, $f$ is a variable over a function sort,
$[ \mkern2mu ]$ is the empty list, and $\_ \triangleright \_$ is the
list prefix operator. Observe that a typical phenomenon pops up,
namely, that function symbols can either be used stand-alone but also
be applied to subexpressions. E.g., a function
$g \colon D \rightarrow D$ can be used as in
$\mathit{maplist}(g,[ \mkern1mu ])$ as well as in $g(g(d_0))$
with~$d_0 \mkern1mu {:} \mkern1mu D$.

As quantifiers are very helpful for concise data descriptions, these
were added to the data language as well. Fermat's last theorem,
\added{$\forall \mkern1mu n,a,b,c \colon \Nat^+ \mkern-1mu .
  \mkern1mu (n>2) \rightarrow a^n+b^{\mkern1mu n} \neq c^n$}, with
$\Nat^+$ the positive numbers, is a perfect expression of type bool in
mCRL2\@. Quantifiers make mCRL2 particularly expressive. The extension
was in line with the general philosophy of mCRL2 that expressiveness
has priority over effective calculability.
\added{Also, lambda
  abstraction was added.} In mCRL2 it is for instance
possible to write $\lambda \mkern1mu n {:} \Nat. \mkern1mu n{+}1$ for
a function that increments a number. Since all expressions in mCRL2
are simply typed, beta-reduction for lambda expressions is guaranteed
to terminate.

The equations were extended to allow conditions. For a conditional
equation to be applicable, the condition must be \added{equal to
  true}. A typical example of this is
\begin{mcrl2}
map&\mathit{fac} : \Int \rightarrow \Nat\,;\\
var&n:\Int\,;\\  
\textbf{eqn}&\mathit{fac}(0)=1\,;\\
            &(n>0)\rightarrow \mathit{fac}(n)=\mathit{fac}(n-1)*n\,;
\end{mcrl2}
Here we use integers ($\Int$) as the domain of $\mathit{fac}$. So, the
minus operator is defined.
The semantics of the data types remained essentially the same as that
for the data types of $\mu$CRL, modulo the extensions mentioned. In
particular the use of function types required so-called applicative
structures in the semantics~\cite{DBLP:books/mit/GrooteM2014}. But in
essence two terms are the same if they are the same in all models of
the data types, or equivalently, if their equality can be proven using
equational reasoning, induction on the constructors, inequality of
true and false, and the use of the standard rules pertaining to
quantification and lambda abstraction.

Inspired by functional programming~\cite{BW88:ph, Has98:report} a
shorthand was introduced to define data types using the keyword
$\textbf{struct}$. For example,
\begin{mcrl2}
  sort&\mathit{PairUndef}=\textbf{struct}~\mathit{undefined} \mid
  \textbf{pair}(\mathit{first}:\Nat,\mathit{second}:\Nat)?\mathit{is\_pair}\,;
\end{mcrl2}
\added{This defines a sort $\mathit{PairUndef}$ consisting of pairs of
  natural numbers and a separate constant $\mathit{undefined}$. The
  optional recogniser function $\mathit{is\_pair}$ is used to
  determine whether a term of sort $\mathit{PairUndef}$ consists of a
  pair. The optional projection functions $\mathit{first}$ and
  $\mathit{second}$ provide the first and second argument of a pair.}

Such struct types can be used recursively, using which, for instance,
list and tree data types can be constructed. \added{Struct types can
  easily be defined using} constructors, mappings and equations, but
in this form \added{they} are far shorter and better readable.

In the same vein of struct data types, it is also possible to define a
name for another, possibly compound type. This is called a sort
alias. E.g., an array can be seen as a function from natural numbers
to some sort~$D$.
\begin{mcrl2}
  sort&\mathit{Array}=\Nat\rightarrow D\,;
\end{mcrl2}

The largest change regarding the data types was the introduction of
standard data types. The basic motivation was to avoid the definition
of the basic data types in each specification over and over again. As
a second benefit it was considered that in every specification the
basic data types are the same. In $\mu$CRL it could happen that two
specifications of the natural numbers differed slightly, but such
differences could easily be overlooked. As a third benefit, the change
allowed to write elements of the data type in the commonly accepted
form. Writing~$3$ in a specification is preferred over
$\mathit{succ} \mkern1mu (\mathit{succ} \mkern1mu (\mathit{succ}
\mkern1mu (\mathit{zero})))$.  The standard data types are defined in
a straightforward fashion (see Appendix~B
of~\cite{DBLP:books/mit/GrooteM2014}). However, in some cases the
definition is quite involved, for instance for the modulo operator or
the integer square root.

Initially, the standard data types were $\Bool$, $\Pos$, $\Nat$,
$\Int$, $\Real$, $\textit{List}(D)$, $\textit{Set}(D)$,
and~$\textit{Bag}(D)$. In the {ASCII} syntax of the tool these were
written as $\texttt{Bool}$, $\texttt{Pos}$, $\texttt{Nat}$,
$\texttt{Real}$, $\texttt{List}(D)$, $\texttt{Set}(D)$,
and~$\texttt{Bag}(D)$. In this text we prefer to use the non-ASCII
notation. Later, finite sets, $\textit{FSet}(D)$, and finite bags,
$\textit{FBag}(D)$, were added. Although an attempt was made to keep
the definition of these standard data types stable, the rules of the
data types have been changed through time. This happened in some cases
to repair errors in the basic definition and in others to make the
rules more efficient, especially regarding its term rewriting
behaviour. \added{Also,} equations
\added{were added} to allow smoother analysis of mCRL2
specifications.

The definition of the standard sort booleans $\Bool$ is
straightforward. The \added{first} major revision was that \added{true
  and false are now written as $\true$ and $\false$, instead of $T$
  and $F$.} \added{Moreover,} the boolean operators could now be used
in infix notation. The heavy-handed $\mu$CRL expression
$\mathit{and} \mkern1mu (x,\textit{T})$ is written in mCRL2 as
$x \land \true$, or in the ASCII format as
\texttt{x\:\&\hspace{-1pt}\&\:true}.
For the standard natural numbers it was considered unwise to define
them Peano-style, using zero and successor, as this is inefficient
both in representation and for calculations. So, it was decided to use
binary notation. For convenience, first positive numbers were defined
using constructors $\mathit{c1}$ and~$\mathit{cDub}(b,p)$ with~$b$ a
boolean and $p$ a positive number. The constructor~$\mathit{c1}$
represented~$1$ and $\mathit{cDub}(b,p)$ stood for $2p$ or~$2p+1$
depending on the boolean. Natural numbers consisted of positive
numbers extended with an explicit extra constructor~$\mathit{c0}$
representing~$0$.

The parser and pretty printer allow to write numbers in the common
standard mathematical way. A tricky set of typing rules determines
whether a number is of sort $\Nat$, $\Pos$, $\Int$, or~$\Real$. If
necessary, implicit type conversions are applied, which can sometimes
be confusing. For instance, in the multiplication $n{\cdot}m$ with
$n:\Nat$ and $m:\Pos$, a type conversion to~$\Nat$ is applied to~$m$,
as multiplication requires both sides to have the same type.  Each
explicit number is then translated into the format of its respective
sort and rewritten using the equations defined for that sort.  The
same applies to the standard operations on numbers like $+$, $*$, $-$,
etc. When using mCRL2 the use of numbers is in general smooth and it
is easy to forget that an extensive definition lies behind it.

Because of the way they are defined, numbers are not bounded and
coincide exactly with the mathematical notions of positive and natural
numbers. The computational complexity is quite agreeable, but
operations on numbers are still substantially slower than machine
operations. In order to employ the advantages of 64-bit machine
calculations, the sorts $\Nat$ and~$\Pos$ were redefined to use
64-bit machine digits in~2023. Any number smaller than~$2^{64}$ 
is represented by a single digit. Larger numbers are represented by
multiple digits. As a consequence, the analysis of a specification
that employs numbers to a large extent is sped up. But this change was
completely internal and most users will not have noticed.
The integers are defined as either a natural number or a positive
number, \added{where the latter is} interpreted as a negative
value.
The reals are essentially just rational numbers, i.e., an integer
divided by a positive number. Extensive attempts to come to a more
general abstract data type of real numbers, which would include
logarithms and trigonometric functions, for instance on the framework
found in~\cite{Pot1735:klr}, were not successful.

Since lists are commonly used it was decided to add a sort
constructor $\textit{List}(D)$, defining the sort of lists of
sort~$D$. Standard notation for lists, as in $[ \mkern2mu ]$ for the
empty list and $[3,4]$ for a list with numbers, was also introduced,
together with some standard infix notation, like pre- and postfixing
an element to a list, or concatenating lists.

Similarly to lists, sets and bags over a domain were incorporated.  It
was considered essential that sets and bags coincide with their
mathematical counterparts. We outline the situation for sets here, but
the story for bags is similar. Sets are written \added{as}
$\{ \mkern2mu\}$ \added{and} $\{1,2\}$, for example, but it is also
allowed to apply set comprehension, for instance,
$\{ \, n\mkern2mu{:}\mkern2mu\Nat \mid {n \mkern1mu |_{\mkern1mu 2}}
\approx 0 \,\}$, where $|_2$ stands for modulo~2, for the set of all
even numbers. Initially, sets were represented as their characteristic
function, in this case a function of sort $\Nat \rightarrow
\Bool$. However, this did not work well when in behavioural
specifications sets were used to add and remove elements dynamically,
leading to complex lambda expressions which were the accumulated
result of additions and removals. Therefore, sets (and bags) are
currently represented as a pair of a characteristic function and an
exception set, which is a finite set of elements for which the set
membership as indicated by the characteristic function has to be
reversed. This turned out to be a very practical and generic
solution. It worked well in all kinds of models, for instance in
models of file systems with sets of open and closed files. Yet, as the
internal structure is relatively complex and most sets are
intrinsically finite, finite sets and bags were added also, internally
consisting of an ordered list.

The problem that occurred initially with sets, namely that insertions
and deletions led to complex lambda expressions, also came about with
functions. A variable of a function sort is a convenient way to
\added{represent} for instance an array. Updating the array with
new values can easily be expressed with a lambda. But the complex and
non-normalisable expressions that result after a number of updates
made this impossible to use for state space exploration.
Therefore, an explicit function update notation
$f[ \mkern1mu n \rightarrow m]$ was introduced which stands for the
function~$f$ except that for argument~$n$ it must yield~$m$.  The use
of functions with changing values in behavioural specifications is now
pleasantly supported.

The second substantial change, going from $\mu$CRL to mCRL2, is
related to communication. To improve communication the notion of a
multi-action $a_1 \mid \cdots \mid a_n$ was introduced. A multi-action
is a super-action where all actions mentioned in the multi-action
happen exactly at the same time~\cite{Arn94:ph}. The multi-action
operator~`$\mid$' is commutative, associative, and has the
action~$\tau$ as unit element. Instead of using a global \textbf{comm}
declaration, a local communication operator $\Gamma_C$ was proposed
where $C$~is a set of elements of the shape
$a_1 \mid \cdots \mid a_n \rightarrow a$ expressing that the actions
$a_1, \ldots, a_n$ in a multi-action can be replaced by a single
action~$a$, \added{expressing} that these actions
synchronise to~$a$.

Because the actions in parallel processes in mCRL2 can happen by
themselves or happen at the same time with other actions, the number
of the multi-actions grows exponentially in the number of parallel
components. Therefore a stronger operator than the encapsulation or
blocking operator~$\partial_H$ was necessary to weed these out. This
motivated the introduction of the allow operator $\nabla_V$ with~$V$ a
set of the only permitted multi-actions. The typical
form of a set of communicating processes in mCRL2 is
$\nabla_V(\Gamma_C(p_1 \parallel \cdots \parallel p_n))$ instead of
$\partial_H(p_1 \parallel \cdots \parallel p_n)$ in $\mu$CRL.

The then-if-else $p \triangleleft c\triangleright q$ of $\mu$CRL was
abolished and replaced in mCRL2 by the if-then-else
$c \rightarrow p \diamond q$, which can also be written as if-then
$c \rightarrow p$.

Consecutive timed actions must happen at strictly later times. So, in
$a\at{1} {\cdot} b \mkern1mu \at{2}$ the action~$b$ can be performed,
but in $a\at{1} {\cdot} b \mkern1mu \at{1}$ the~$b$ cannot be done as
it must happen strictly after~$a$. This resembles the semantics in
\cite{DBLP:journals/fac/BaetenB91}. As mentioned, the latter
expression is equal to $a\at{1} {\cdot} \delta\at{1}$ with a time
deadlock at time~$1$ indicating that time cannot proceed beyond
time~$1$ without violating the prescribed behaviour in the process
specification. With the use of multi-actions it is possible to express
that actions $a$ and~$b$ happen at the same time, i.e.,
$(a\mid b)\at{1}$ in a much more natural way.

Besides the difference with $\mu$CRL discussed, the process part in
mCRL2 was in the course of time provided with two more extensions. The
first one was to allow assignments at the right-hand side of process
equations, e.g.,
\begin{mcrl2}
  act & a;\\
  proc&P(x,y:\Nat) = a{\cdot}P(x=x+1);
\end{mcrl2}
This indicates that parameter~$x$ will be increased, whereas
$y$~remains unchanged, leading to~$P(x+1,y)$ where the update of
parameters is based on their position. It is noted that in case
a process has a large number of parameters, positional updates are
quite unreadable, especially when most parameters do not change value.

A second eventual change to mCRL2 processes was the extension of the
language with probabilities, as probabilistic behaviour is very
intriguing and crucial to the modelling of some systems
\cite{DBLP:conf/birthday/GrooteW25,
  DBLP:journals/corr/abs-2407-06809}.  In short, it is now possible to
write $\textbf{dist}~d{:}D \mkern1mu [f(d)]. \mkern1mu p(d)$ which
expresses that process~$p(d)$ is executed where~$d$ of sort~$D$ is
chosen according to distribution~$f$. Although this works nicely,
there are still many fundamental questions to be answered
around the interaction of probabilities and nondeterminism, especially
with uncountable nondeterminism.

\section{Modal logics}

The initial idea of analysing systems in $\mu$CRL was to model an
implementation and a specification, hide internal actions in the
implementation, and prove the implementation equal to the
specification with respect to weak or branching bisimilarity
\cite{DBLP:books/sp/Milner80,
  DBLP:journals/jacm/GlabbeekW96}. However, this turned out to be
cumbersome as it is generally difficult to write specifications that
are behaviourally the same as implementations. \added{Allowing more
  relaxed behavioural relations did not really resolve this issue
  \cite{Gla93:concur, DBLP:series/txcs/Roscoe10}.}  Back then, often
the most fruitful approach was to generate a state space, hide many
actions, and view the result modulo \added{some weak behavioural
  equivalence}. With a bit of luck the resulting state space was small
and insightful and would therefore provide the required understanding
of the model.

In general, however, to effectively analyse behaviour, some property
language was required. \added{These are added to virtually all
  specification languages, with CSP as notable exception
  \cite{DBLP:books/ph/Hoare85,Ros97:ph} where properties are encoded
  in processes and verified using failure preorders.} Modal languages
such as LTL, CTL, and~CTL* \cite{Pnu77:focs, EC82:scp, EH82:stoc,
  Eme90:handbook, BK08:mit} are less in line with process calculi
based on actions. Hennessy-Milner logic with its diamond and box
modalities, $\langle a \rangle \varphi$ and $[a]\varphi$, was more
fitting~\cite{HM80:icalp,HM85:jacm}. But as Hennessy-Milner logic is
essentially restricted to finite behaviour, it had to be extended with
recursive features, for instance provided by the fixed point
modalities in the modal mu-calculus~\cite{Koz83:tcs, Sti01:springer,
  BS07:handbook,BW18:handbook}. As fixed point modalities were
considered difficult, multiple attempts were made to come up with a
`better' logic, see e.g., \cite{GrVl95}.  But the elegance and
expressivity of the modal mu-calculus turned out hard to beat,
leading to its adoption. Within the context of $\mu$CRL this was
mainly experimental, but in mCRL2 modal formulas now form an
indispensable part.

The modal mu-calculus provides minimal and maximal fixed point
operators, respectively $\mu X.\phi$ and $\nu X.\phi$ where $\phi$ is
a Hennessy-Milner formula that can also contain $X$, as well as other
fixed point operators. In these formulas the data types of mCRL2 can
be used, including quantifiers and lambda expressions. The maximal
fixed point operators are used to formulate safety properties, minimal
fixed point operators are used for liveness properties, and by mixing
minimal and maximal fixed point operators fairness properties can be
expressed.

There is one extension which is very important to express complex
requirements in this logic. This is the addition of parameters to the
variables in fixed point operators. These parameters are often
referred to as observation variables, because they can be viewed as
the variables into which an observer accumulates its observations. For
example, the following formula expresses that the number of actions
$\mathit{up}$ cannot exceed the number of actions $\mathit{down}$. The
observation variable~$n$ records the difference between the numbers of
the two actions.
\begin{mcrl2}
  form & \nu X(n:\Int=0).[\mathit{up}]X(n+1) \land
  [\mathit{down}](n>0\wedge X(n-1)) \,;
\end{mcrl2}
Initially, it was expected that modal formulas were intrinsically
small. But this turned out not to be the case. Formulas of hundreds of
lines with a dozen or so parameters do occur regularly when analysing
actual systems. In this form the modal mu-calculus is superior to
CTL*. Any CTL* formula can be linearly translated to the modal
mu-calculus, while the algorithms to solve them are of the same
complexity \cite{DBLP:journals/tcs/CranenGR11}. Reversily, there are
many properties not expressible in CTL* which can be expressed
naturally in the modal mu-calculus. \added{These comprise not only
  formulas with substantial book-keeping of data parameters, but also}
various variants of fairness properties.
In the timed setting, when extended with an until-like operator the
mu-calculus is more expressive than~TCTL, see~\cite{CKF24:qest}.
For determining whether a modal formula holds for
an mCRL2 process, Parameterised Boolean Equation Systems have been
developed. More about them can be found in
Section~\ref{sec:modelchecking}.

\section{Behavioural analysis}
\label{sec:verif}

The conciseness of $\mu$CRL made it an attractive vehicle to show the
correctness of behaviour.  Process algebra per se was less suitable,
as it handles data rather haphazardly, mainly as indices of process
variables. In $\mu$CRL data \added{is} a first-class citizen, and
with the defining equations, the properties of the data are
unambiguously defined.

Analysis with $\mu$CRL went through a number of stages. Initially, the
behavioural analysis consisted of manually proving implementations of
distributed algorithms and protocols equal to their specification,
culminating in the \added{so-called} cones and foci method
\added{(see below)} and the correctness proof of Tanenbaum's third
sliding window protocol. But later it was realised that most practical
specifications would be so large, that manual verification was not the
way to go. At that time, tools focussed on state space generation and
reduction. Later, visualisation and deciding the validity of modal
formulas came into play. Currently, the question is how to effectively
deal with large systems, not only by improving verification
technology, but also by learning how to model systems such that they
can be verified~\cite{DBLP:journals/stvr/GrooteKO15}.

\subsection{Manual verification}

As indicated, $\mu$CRL is a very compact language, of which the
constituents came with a well developed mathematical theory. The
mathematical elegance was combined in a proof system for
$\mu$CRL~\cite{GP93:sosl}, which in a slightly smoother shape can
also be found in~\cite{DBLP:books/mit/GrooteM2014}. 
\added{This proof theory consists of the axioms for process algebra,
the equations and induction principles from the data definitions,
the Recursive Specification Principle (RSP) to prove recursive
specifications equal to each other, and Koomen's Fair Abstraction 
Rule, although the latter is hardly used. }

Basically, the question was whether we could prove implementations of
distributed systems equal to a separately formulated
specification. Equality was generally taken to be branching
bisimilarity. It was believed that for practical systems there is no
difference between branching bisimilarity and weak bisimilarity and
that one could use either. Later it turned out that this is not
\added{exactly} true. For instance, Peterson's mutual exclusion
algorithm is slightly different when reduced modulo branching
bisimilarity or weak bisimilarity.

A challenge formulated at the end of the eighties was whether the third
and most complex sliding window protocol in Tanenbaum's textbook
on computer networks could be proven correct~\cite{Tan81:ph}. The
simpler versions of sliding window protocols in the book were
investigated and proven correct after being
repaired~\cite{Vaa86:msc}. Two attempts to prove the correctness of
the last variant of the protocol were not
successful~\cite{Gro87:msc,Mid86:msc}.
We embarked on this task, but it became clear that it was a much
longer journey than anticipated, with the final result published in
\cite{FGPBP04:amast}, 14 years after starting to look at it.  A
shorter version of the proof is available
in~\cite{DBLP:books/mit/GrooteM2014}.
We describe here the phases we went through. Each phase added a new,
and as we believe essential, proof methodology.

It started out by understanding the third sliding window protocol.
After approximately one month it became clear that modelling the
protocol gave rise to two problems. The first problem is that its
external behaviour is very hard to formulate, as the protocol delivers
data in bursts. All data that has been properly transferred has to be
delivered first to the receiver, before data can be accepted from the
receiver to be sent to the other side. The precise form of the bursts
is determined by the pattern of the data lost in transit.

The second issue was even worse. There is a command
\textit{StopAckTimer} with a comment ``no need for separate ack
frame'' \cite[page 162, fourth Indian edition]{Tan81:ph}, which would
cause a livelock as it may give rise to a situation where data being
resent from one side to the other will never be acknowledged and
therefore, no new data will ever be transferred. In the propositions
of~\cite{Groote91} this problem is mentioned. Later editions of
\cite{Tan81:ph} are free of this bug. At the time the problem was not
known to Tanenbaum. He actually even claimed that it could not be
important, as during a decade of implementations, it went unnoticed.
This is because the bug has a small probability to take place. It only
occurs after transmitting at least one `window' full of data while
losing an acknowledgement at exactly the right moment. The livelock
is broken when data is sent in the reverse direction, as piggy backed
acknowledgements can release the livelock. But with no traffic in the
reverse direction, this third sliding window protocol can stop
transmitting data at arbitrary moments, and with scarce traffic in the
reverse direction, it typically can stall so now and then for no
apparent reason.

\added{With the gained insight} it is not particularly difficult to
rephrase the sliding window protocol such that it does not have the
flaw mentioned above and behaves as a bounded queue twice the size of
the window used in the protocol, see e.g.,
\cite{DBLP:books/mit/GrooteM2014,FGPBP04:amast}. Proving that both are
branching bisimilar is another matter. The first attempt was to use
commutativity and associativity of the parallel composition, by
splitting the sliding window in multiple alternating bit protocols.
This did not work and led to the paper~\cite{GK94}, which incidentally
also contained a protocol which was \textit{not} the well-known
Lamport's bakery protocol~\cite{Lam74:cacm, Lam86a:jacm}. Together
with the results in~\cite{DBLP:conf/concur/GrooteM92}, where it was
also tried to prove equality of systems based on the properties of the
parallel composition, it was concluded that this led to a dead end.

All subsequent approaches would start to eliminate the parallel
composition by first transforming the implementation and the
specification to a linear process,
cf.~\cite{DBLP:books/mit/GrooteM2014}. Basically, a linear process is
a process where the state is represented by the parameters of the
process. Using conditions, actions, and next state descriptions it is
indicated when actions can be done and what their effect on the state
is. Every guarded recursive process can be transformed to a linear
process~\cite{Use02L:phd}, generally without notable growth. A linear
process is therefore a convenient normal form which not only became
the basis for verification, but also became the internal format for the
tools.

The task from this point on was to prove that the implementation and
the specification as linear processes are equal. It was observed that
invariants were most likely needed. Invariants are the workhorse in
proving the correctness of sequential programs and were also the core
behind sequential program derivation techniques \cite{Kal90:ph,
  FG99:springer}. But in process algebra, they were non-existent. For
the sliding window protocol invariants are very convenient to
formalise how the contents of the receiving and sending message
buffers relate. In~\cite{DBLP:conf/concur/BezemG94} a special rule was
formulated, namely CL-RSP with invariants, which allows to prove that
two linear processes were equal given an invariant. Remarkably, it
could be shown that this rule is derivable from CL-RSP, RSP for
convergent linear processes, which allows to show that two linear
processes are equal, and which does not explicitly use
invariants. This shows that invariants are theoretically
superfluous. For practical verification they are indispensable.
Although some smaller examples could be proven correct with elegant
and straightforward proofs, CL-RSP with invariants was not enough to
prove the correctness of Tanenbaum's third sliding window protocol.

One of the parts that blocked finding a proof was the following.  If a
step was done in the specification, for instance the delivery of some
data element that was successfully transferred by the protocol, it
could take an unbounded number of steps to mimic such a step in the
implementation. In the implementation the data element \added{may} not
have been transferred yet to the receiving side, and while attempting
to be delivered could be lost for an arbitrary number of times. This
would mean that mimicking a step from the specification in the
implementation would require a long and tedious proof, making the
overall proof long and unattractive.  The cones and foci method was
designed to deal with exactly this
situation~\cite{DBLP:journals/jlp/GrooteS01}. It was substantially
improved in \cite{DBLP:journals/entcs/FokkinkP05a,
  DBLP:journals/fmsd/FokkinkPP06}. In essence the method allows to
show branching bisimilarity between an implementation and a
specification by only showing that the implementation can mimic the
actions in the specification at so-called focus points. Furthermore,
it needs to be shown that the implementation will always move towards
the focus point. The states leading up to a focus point are called the
cone. Concretely, the method yields six proof requirements from which
branching bisimilarity will follow. With the use of invariants, these
six requirements are in general not too hard to prove.

Using the cones and foci method compact proofs of various and highly
nondeterministic protocols could be given
\cite{DBLP:journals/tcs/FredlundGK97, DBLP:journals/fac/GrooteMS05,
  DBLP:conf/fsen/Spaendonck23}. Unfortunately, the cones and foci
method did not work to prove Tanenbaum's third sliding window
protocol. The reason is that it uses sequence numbers modulo twice the
buffer size in each protocol entity.  The invariants describing which
ranges of sequence numbers contain valid transmitted elements could
not be formulated when working in modulo arithmetic. The solution was
to first prove a variant of the sliding window protocol with an
infinite range of sequence numbers using the cones and foci
method. Following that, the implementations working with and without
modulo sequence numbering were proven equal~\cite{FGPBP04:amast}. This
technique was also applied by Schoone
using assertional verification~\cite{Dij76:ph, Lam77:tase, Sch96:cup}.

Further development of manual proof methods came to a halt as
attention switched to developing automated proof tools. A typical
question which is still open is to come up with proof methods for
other equivalences than branching bisimilarity, for instance along the
lines of~\cite{AL91:tcs, Lam02:aw} where trace-like equivalence is
established using so-called prophecy variables. These techniques are
for instance needed to give concise correctness proofs for atomic
registers of which the implementation is generally not bisimilar to
the specification~\cite{Lam86a:jacm, Lam86b:jacm, Hes98:ipl}.

\subsection{Toolset}

The desire to show that the process algebraic analysis methods are
practically relevant led to numerous contacts, a.o., with Philips
Natlab.  A large project upon which we embarked in the mid nineties
was to model the software and the protocols of the KidCom, later
renamed to Intuit. This was a handheld device with a touch screen and
infrared communication. The device would allow for wireless
communication, albeit one ought to be in the same room, and would
allow to exchange text and pictures, and play various sorts of
games. It was even foreseen to have options to exchange hidden
messages that could be displayed later, for instance for
advertisement. We made an extensive model of the behaviour of the
KidCom in PSF~\cite{KidCom} alongside the development of the software.
This document contained 169~pages, and it was used for model based
testing the infrared communication software of the KidCom.

Two conclusions were drawn from this project. The first one was that
formal models are not necessarily correct. While testing the infrared
communication on the basis of the model, it was detected that the
model contained deadlocks and was clearly flawed. Although it is
obvious that this can happen, this came somewhat as a surprise then as
it was silently assumed that when writing down formal models these
would by definition be correct. Through time it was realised that
correctness does not come by formal descriptions themselves, but, by
redundant views that can be compared~\cite{DBLP:conf/ifip/Brooks86,
  DBLP:journals/tse/XieE03, DBLP:journals/scp/BrandG15}. This is one
of the primary reasons that it is wise to analyse behavioural models
by separately formulating and checking modal properties.

The second conclusion was that it is virtually impossible to prove the
correctness of the models of the behaviour of the KidCom by hand. And
even if this would be possible by experts using extensive effort,
software developers would not easily acquire the skill and may miss
the time or perseverance to bring the analysis to a successful end. If
we wanted formal techniques to be applied at scale, more accessible
analysis needed to be available. This was the primary motivation for
starting the development of a toolset for $\mu$CRL. 

\added{At that time a few other verification tools were available.
  Around LOTOS a large project called LOTOSphere had led to a series
  of tools, including simulators and generators of extended finite
  state machines \cite{lotosphere}. The only surviving part of this
  work appears to be the CADP toolset \cite{CADP96:cav}. Around CSP
  the FDR (Failure Divergence Refinement) tool was
  built~\cite{DBLP:conf/spin/BroadfootR00}.  The language PSF (Process
  Specification Formalism) based on ACP had been provided with a
  simulator \cite{DBLP:journals/fuin/Veltink10} and elementary state
  space generation facilities. The most accessible and widely used
  tool at the time was SPIN around the specification language Promela
  \cite{DBLP:journals/tse/Holzmann97, DBLP:books/daglib/0020982},
  which in essence was an elementary C-like language with a minimal
  but adequate set of data types, instead of a process algebra. The
  tools CADP, FDR, and SPIN are still actively maintained and publicly
  available, although CADP and FDR can only be used freely for
  education and research.  As $\mu$CRL as a language was established
  and different from the other languages, it seemed easier to built
  tools from scratch than adapting existing software.  Directly from
  the outset, it had been determined that the $\mu$CRL and mCRL2 tools
  should also be free to use in commercial settings, via the Boost
  licence, as its development is largely publicly funded and its use
  should be stimulated.}

A secondary motivation came from the observation that many research
groups in computer science have little or no experience with the
development of software at large.  In order to understand what are the
true problems of the development and maintenance of real software, one
has to start working on a large and serious software project.

The toolset was initially developed for $\mu$CRL. With the advent of
mCRL2 the toolset for $\mu$CRL was used as the basis for an mCRL2
toolset. For a while both toolsets lived alongside each other, being
maintained by independent groups. $\mu$CRL was still maintained at
CWI, whereas the development of mCRL2 took place in Eindhoven. The
combined effort spans a period of more than 30 years now. We sketch
the major developments that occurred during this period.

\subsection{ToolBus and ATerms}

The first version of the toolset was based on the ToolBus architecture
developed by Klint and
co-workers~\cite{DBLP:journals/scp/BergstraK98} to interconnect
various components. The idea was that a toolbus architecture leads to
independence of the components as they communicate via the toolbus
over well-defined interfaces. This should give rise to better
maintainable software. This first toolset is completely formally
specified in~\cite{DG95}. It also contains its complete implementation
in~C.

The ToolBus uses so-called ATerms to exchange information between the
different components~\cite{DBLP:journals/infsof/BrandK07}. An ATerm is
in essence just a term, i.e., a function symbol applied to
subterms. In various variants, terms can have other shapes, such as
lists, numbers, and even \added{blobs}, arbitrary blocks
of data.  The~A in ATerm stands for annotated, as there are means to
attach metadata to an ATerm. We have never used this possibility.
ATerms are stored with maximal sharing. This means that each term
occurs at most once in memory per ATerm repository. Furthermore,
ATerms are automatically garbage collected, a feature not available
in~C. Due to the sharing of ATerms, subterms cannot be changed
directly. This has the disadvantage that changing a term means that
the whole term must be copied first before modification. \added{A}
significant advantage is that possibly sizeable terms can be passed
along by just forwarding a single pointer \added{and determining
  whether such terms are equal only requires one pointer
  comparison}. 

The ToolBus exchanged ATerms between the various
components. Unfortunately, all components and also the ToolBus itself
had their own ATerm repositories. \added{This implied that in case of
  transferring a term from one component via the ToolBus to another
  component,} this term was copied at least two times. Although
efficient implementations were conceivable by letting all components
use the same repository and only passing pointers along, this never
materialised. The ToolBus became an unbearable performance bottleneck
and was therefore abolished. ATerms still form the core of the toolset
up to this day.

At the end of the nineties CWI in Amsterdam acquired a large
32~processor, 64~Gigabyte, 64-bit SGI computer for large-scale
calculations, called Medusa.  This led to the desire to transform the
32-bit ATerm library written in~C to 64-bit. Unfortunately, the
process took multiple years. The two reasons for this were the
following. First, research groups do not have the culture of giving
high priority to servicing the users of the research software. Second,
the C-implementation of ATerms was highly convoluted, with various ad
hoc bit-level optimisations, which made the adaption to 64-bit a
non-trivial endeavour. In the meantime the rather expensive Medusa
stood mainly idle, only being used for factorisation of large numbers,
which could far more \added{cost-}effectively be done on a set of
ordinary desktops~\cite{AGLL94:asiacrypt}. The research group behind
the ToolBus and the ATerm library switched to the use of Java. This
meant that -- with explicit permission -- we started to maintain a
copy of the C~ATerm library ourselves. The library has been
reimplemented in a straightforward manner in~C++, using only the most
essential datatypes for the toolset, namely, terms, lists and numbers,
and removing features like \added{blobs} and
annotations. It still heavily relies on garbage collection and maximal
sharing of terms.

The existence of the multiprocessor machine Medusa sparked the desire
to come up with a parallel ATerm library such that ATerms could be
constructed and used simultaneously. A first ad hoc attempt turned out
to be faulty, due to the ABA problem \cite{IBM83:manual, Mic04:tpds}.
This problem says that when inserting and deleting elements from a
shared list using the compare-and-swap instruction, the list can
become completely mixed up, although this only happens under very
specific circumstances. This is another nice example that shows that
testing of software is inadequate to detect errors, and taught us that
one cannot trust even the simplest algorithms without proper
correctness analysis. A proposal to implement basic ATerms correctly
can be found in \cite{DBLP:journals/dc/HesselinkG01}.
However, this basic algorithm did not allow for sharing and parallel
garbage collection of terms. As it was believed at the time that wait-
and lock-free algorithms were necessary to obtain good performance,
both an almost wait-free parallel hash table and a lock-free parallel
garbage collection were defined in \cite{DBLP:journals/dc/GaoGH05,
  DBLP:journals/scp/GaoGH07}. In order to be sure that these
algorithms were correct, the Herculean task was undertaken to prove
the correctness of the algorithms in PVS~\cite{ORS92:case}, which took
multiple years. Prototype implementations turned out to be correct,
but the performance of these algorithms lagged behind, in the sense
that single processor solutions generally outperformed multiprocessor
implementations.

Many years later a new attempt was done to make a parallel
implementation. This time the focus was on avoiding contention,
\added{while} allowing the use of explicit synchronisation. A
special busy-forbidden algorithm was designed employing the way terms
are being accessed in ATerms. At first, the algorithm was proven
correct using model checking for a restricted number of parallel
processes.  Later, it was proven correct for an arbitrary number of
processes using the cones and foci method
\cite{DBLP:conf/fsen/Spaendonck23}. The new parallel implementation of
the ATerm library scales almost linearly with the number of processors
on adequate processor architectures \cite{DBLP:conf/isola/GrooteLS22,
  DBLP:conf/birthday/GrooteJLSW22}.  One of the conclusions we drew
from this is that model checking is far more \added{efficient} in
designing correct protocols than manual correctness proofs, especially
if they are also computer checked.

\subsection{Term rewriting}

The workhorses behind equational data types are term rewriting
engines.  The first implementation of the toolset used a
straightforward but slow innermost term rewriting system~\cite{DG95}.
By some optimisations the performance became reasonable. The most
effective optimisation was to combine substitutions with rewriting,
always storing normalised terms in substitutions. \added{As a
  consequence when rewriting a term with variables, substituents for
  these variables themselves never needed to be normalised.}

Innermost rewriting sometimes rewrites unnecessarily. This for
instance applies to if-then-else expressions such as
$\textit{if} \mkern2mu (c,t_1,t_2)$.  First, the condition~$c$ and
both terms $t_1$ and~$t_2$ are rewritten to normal form. Subsequently,
depending on the value of~$c$ either $t_1$ or~$t_2$ is chosen. It is
clear that too much work is done, and that only one of the two terms
needs to be rewritten. For this problem just-in-time rewriting was
invented \cite{DBLP:journals/tcs/Pol01,DBLP:conf/rta/Pol02}. Inspired
by \cite{DBLP:journals/toplas/BrandHKO02} a compiling rewriter was
developed that translates a rewriting system initially to C and later
to C++ \cite{DBLP:journals/entcs/Weerdenburg07, Wee09:phd}.
In~\cite{GTA18:wrla} an extensive comparison of term rewriters was
performed. This showed that the performance of the mCRL2 rewriters can
still be improved. An extensive study has been performed to increase
the speed of rewriting by optimising the matching of rules using
automata and by avoiding unnecessary constructions of terms during the
rewrite process~\cite{Erk24:phd}. These algorithms are not yet part of
the toolset.

\subsection{Linearising processes}

The manual verification of process equivalence suggested that it is
advantageous to transform each process to linear form. As mentioned, a
linear process has its full state encoded in data and is essentially
defined by a set of condition-action-effect rules. The linear process
is a central notion in the verification approach, both manual \added{and}
tool-supported.

For the toolset every $\mu$CRL process and later also each mCRL2
process is first transformed to linear form before any other operation
is applied to it. The tool achieving this is called the
lineariser. First attempts to build a lineariser were made by Bosscher
in ASF-SDF \added{(Algebraic Specification Formalism--Syntax
  Definition Formalism)}~\cite{BP95:tacas}, improved by Luttik using a
more generic rewriting approach~\cite{LV97:tpas}. The work of the
latter influenced the development of the Stratego workbench
\cite{BKVV08:scp, Vis01:rta}. As both projects did not lead to a
successful lineariser, an implementation has been constructed from
scratch in~C, which, being transformed to C++ is still in use. This
lineariser works on the most commonly used subset of mCRL2, but does
not cover the whole language. For instance, it does not allow to mix
recursion with the parallel operator. Usenko showed that the full
$\mu$CRL language can be linearised~\cite{Use02L:phd}, but this
translation has never been fully implemented.

A linear process is a very pleasant normal form, and therefore all
other tools are based on it. The mCRL2 toolset includes for instance
tools that eliminate non-used parameters or parameters that remain
constant~\cite{DBLP:journals/sigplan/GrooteL02}. Optimising the linear
form by the appropriate tools is an integral part of performing a
behavioural analysis, especially when the behaviour is extensive.
The current notion of a linear process \added{is extended} with
time \added{and} probability distributions.

\subsection{State space generation}

The earliest form of automated analysis with $\mu$CRL was to generate
a state space. The first realistic state space that was generated was
the 74~state state space of the alternating bit protocol \added{with
  two data elements}, which initially took many hours to be
generated\added{, although this was quickly vastly improved}.
Analysis through simulation was also possible, but generally running
simulations gives very little insight in the behaviour of
systems. Using linear processes it is pretty easy to build tools such
as a state space generator. Initially, the initial state, as a vector
of values, is considered unexplored. As long as an unexplored state
remains, all conditions of the condition-action-effect triples of the
linear process are evaluated for this state, yielding a list of
outgoing transitions. The states that are reachable via these
transitions are then divided in unexplored and already visited
states. State vectors are very compactly stored as ATerms, due to term
sharing, in the form of balanced trees. Depending on the model, and
using caching of values satisfying specific conditions, currently more
than $10^5$ states can be generated per second. The essential problem
is that state spaces can become very large, making it impossible to
explicitly generate and store all states~\cite{Cla09:lics}.

At the end of the nineties, computers were generally 32-bits and
carried RAM memories of 4~Gbyte at most. This memory limit was a
serious issue. Therefore, methods were designed to generate a state
space on multiple computers, using the memory of all of them to store
the state space. The trick was to define a hash function that a priori
would already distribute all states over the computers. The essential
bottleneck was the exchange of states between
computers~\cite{DBLP:journals/logcom/BlomLP011}. With the advent of
64-bit machines this approach became outdated.

An other approach sought to reduce the state space before generating
it by means of $\tau$-confluence~\cite{DBLP:journals/tcs/GrooteS96,
  DBLP:conf/mfcs/GrooteP00}, a variant of partial order reduction
\cite{Pel93:cav, God96:lncs}. In essence, $\tau$-confluence states
that if a $\tau$-transition does not cut away options in behaviour,
checked through the detection of confluent diagrams, then all other
transitions in a state where a confluent~$\tau$ can be done, can be
ignored. This can reduce the state space exponentially. Unfortunately,
models are rarely $\tau$-confluent, Milner's scheduler
\cite{DBLP:books/sp/Milner80} or the Dolev, Klawe, and Rodeh leader
election protocol~\cite{DBLP:journals/tcs/FredlundGK97} being notable
exceptions.

\added{Prior to the use of modal formulas and model checking, t}here
were essentially two ways to analyse a generated state space. Simple
properties can be shown, like deadlock freedom or the absense of
certain actions such as an action \textit{illegal}. The second method
is to reduce the state space using behavioural equivalences,
especially branching
bisimilarity~\cite{DBLP:journals/jacm/GlabbeekW96} for which efficient
algorithms exist~\cite{DBLP:conf/icalp/GrooteV90,
  DBLP:conf/tacas/JansenGKW20}. If the resulting state space turns out
to be small, it can subsequently be visually inspected, often
revealing the dynamics of the process. If state spaces are still too
big after reduction, more actions can be hidden.

Using coarser equivalences than strong or branching bisimilarity,
state spaces can be reduced further.  The following equivalences are
at present available in the mCRL2 toolset: weak bisimilarity,
simulation equivalence \cite{DBLP:conf/cav/GlabbeekP08}, strong and
weak trace equivalence, probabilistic strong bisimulation
\cite{DBLP:journals/algorithms/GrooteVV18}, and ready simulation
\cite{DBLP:conf/sac/Gregorio-Rodriguez15}. Besides these there are
other equivalences and preorders that can be used to show two state
spaces equivalent such as strong and weak failure equivalence and
preorder \cite{Ros97:ph} and coupled
similarity~\cite{DBLP:conf/tacas/BispingN19}.

Now, in 2026, state spaces that can be generated explicitly and
reduced can have in the order of $10^{10}$~states. Many models have
more states, and these can be generated using so-called symbolic
methods where the state space is generally encoded in variants of
binary decision diagrams (BDDs)~\cite{BCMDH92:ic}. For mCRL2 the
generic LTSmin toolset is available for some time now to symbolically
generate such state spaces~\cite{DBLP:conf/cav/BlomPW10}. Dedicated
tools for mCRL2, primarily directed to parity games, but also
implemented to symbolically generate state spaces, have been added
more recently~\cite{DBLP:conf/tacas/LaveauxWW22}.
 
\subsection{Model checking}
\label{sec:modelchecking}

As indicated, we decided to use the modal mu-calculus as property
language because of its expressivity. Inspired by~\cite{Mat98:phd} we
started to use parameterised boolean equation systems (PBES) as the
framework for algorithms to calculate whether a formula is valid for a
process~\cite{DBLP:conf/amast/GrooteM98,DBLP:journals/tcs/GrooteW05}. A
PBES is a sequence of boolean fixed point equations where the fixed point
variables can have parameters. For example,
\begin{mcrl2}
  pbes&\mu X(n:\Nat) = (n<3) \lor ( X(n+1) \land Y(n>3) ) \,;\\
  &\nu Y (b:Bool) = X(0) \lor Y(\neg \mkern1mu b)\,;\\
init&X(1)\,; 
\end{mcrl2}
Parameterised boolean equations can easily be obtained from a linear
process and a modal fixed point formula. The formula holds for the
process iff the solution for the initial PBES variable is
true. The PBES framework is very expressive and a wide range of other
problems can be translated to it, including checking equivalence of
processes~\cite{DBLP:conf/concur/ChenPPW07}.

The first model checker for PBESs was based on fixed point
iteration~\cite{DBLP:conf/fmco/GrooteW03} involving BDDs extended with
equality~\cite{DBLP:conf/lpar/GrooteP00}. This worked great in the
sense that modal properties over infinite state spaces could be
proven, but it was not particularly effective for analysing models of
larger industrial systems
like~\cite{DBLP:journals/corr/abs-2403-18723} due the more complex
structure of their state spaces.
Also alternation of fixed points turned out to be problematic.
Hence, the tool \texttt{mucheck} was never transferred from the
$\mu$CRL to the mCRL2 toolset.

A subsequent approach \cite{DPW08:ictac, PWW11:ic} was to transform a
PBES to a boolean equation system (BES), i.e., to a set of boolean
fixed point equations where the variables do not have
parameters~\cite{Mad97:phd}. This process is very comparable to
instantiating a linear process to become an explicit labelled
transition system\added{, where BES variables correspond to states}. 
A boolean equation system is equivalent to a parity
game~\cite{EJ88:focs, Zie98:tcs}. Using the parallel ATerm library it
turned out to be straightforward to implement this instantiation in
parallel~\cite{DBLP:conf/birthday/GrooteJLSW22}. Solving boolean
equation systems is currently done via the recursive
algorithm~\cite{Zie98:tcs, Fri11:rairo} which is the most efficient in
practice \cite{Dij18:tacas, SWW18:gandalf}. Execution time can be
saved by already partially solving a BES when it is being generated,
from which it can possibly be derived that there is no need to
investigate some parts. This is comparable to on-the-fly solving of
modal formulas \cite{CVWY92:fmsd, Hol96:cs}.

A PBES is a structure comparable to a linear process. In a similar way
it can be analysed and optimised by itself. Actually, PBESs are much
more amenable to such optimisations as only the validity of the
initial PBES variable needs to be preserved. Removing unused or
constant parameters, simplifying parameters with complex data types,
simplifying parameters by data flow analysis, applying forms of
partial order or confluence reduction, and even reducing BESs modulo
bisimilarity are some of the applied techniques
\cite{DBLP:journals/tocl/KeirenRW12,
  DBLP:journals/sttt/NeeleWWV22,
  DBLP:conf/atva/KeirenWW14,
  DBLP:journals/tcs/StramagliaKN25,
  DBLP:conf/hvc/KeirenW09,DBLP:conf/cade/Neele22}.

Various other solving techniques have been applied to PBESs, such as a
translation to SMT solving~\cite{DBLP:conf/atva/KoolenWZ15} and
solving PBESs with unbounded data domains by symbolically partitioning
the BES variables~\cite{DBLP:conf/facs2/NeeleWG18}.
The latter technique also allows for real-time verification. Yet,
the most successful approach is to symbolically evaluate PBESs. This
was already done in the tool
LTSmin~\cite{DBLP:conf/tacas/KantLMPBD15}. The current symbolic
solving tools for PBESs can efficiently solve PBESs
\cite{DBLP:conf/tacas/LaveauxWW22} and have been successful on real
systems with well over $10^{50}$ BES variables.

If a modal formula turns out to be invalid, it can be very hard to
figure out why. For this purpose the generation of counter examples,
and dually of witnesses, has been added
\cite{DBLP:conf/cade/WesselinkW18, DBLP:conf/tacas/StramagliaKLW25}.
The outcomes are labelled transition systems that indicate why a
certain formula does or, more interestingly, does not hold. As the
modal mu-calculus is very expressive, the counter examples and
witnesses, collectively called evidence graphs, are not simple traces
or lassos, but can become quite complex. For instance, the formula
expressing that a player has a winning strategy in a game, has as
evidence a tree, precisely indicating which move must be done
\added{to counter each possible move of the opponent}.

For probabilistic processes the fixed point modal formulas have been
extended such that they can also express
probabilities~\cite{DBLP:conf/concur/GrooteW23}. This leads to
(Parameterised) Real Equation Systems (PRESs and~RESs). The toolset
contains some algorithms to solve PRESs, but better algorithms should
be devised. Still, the current framework allows for insightful
analysis of probabilistic behaviour
\cite{DBLP:conf/birthday/GrooteW25,DBLP:journals/corr/abs-2407-06809}.
 
\subsection{Visualisation}

\added{Visualising small transition systems is very insightful
  \cite{DBLP:conf/fm/ValmariS96}.  However, this does not work with a
  larger number of states.} Slightly before the year 2000
investigations started into visualisation of large state spaces,
showing the structure of state spaces of up to~$10^6$ states. A rather
effective, yet simple idea of doing so is reported
in~\cite{DBLP:conf/tacas/GrooteH03, DBLP:journals/sttt/GrooteH06}. The
idea is to transform a labelled transition system into a tree by
removing transitions that go back and grouping all states with a
common substate. This tree can then be visualised easily providing
remarkable insight.

This successful visualisation posed the question whether the
dependency of the parameters in linear processes could be visualised
also. This led to a technique to visualise how those parameters
influence each other~\cite{DBLP:conf/iv/PretoriusW05,
  DBLP:journals/cgf/PretoriusW08}. It was also attempted to visualise
the structure of traces that could be executed by the
system~\cite{DBLP:conf/apvis/PretoriusW08}. Contrary to the other two
visualisation tools, this latter tool is not available anymore in the
mCRL2 toolset. There is another visualisation tool \added{called
  \texttt{ltsgraph}}, which is actually most frequently used. It
depicts transition systems using a force-directed positioning
algorithm and it allows user interaction to place the states. This
tool works comfortably for relatively small labelled transition
systems, but if well-structured, also state spaces of $10^3$ states
can certainly be visualised nicely.

\section{Overview of applications}

Already at the early days of process algebra in the eighties in
Amsterdam, it was important to show that the process algebraic
approach was effective in providing insight in applications,
especially for specifying communication protocols and proving their
correctness \cite{DBLP:conf/aii/BergstraK85, Vaa86:msc,
  wamel1992algebraic, MauwVeltink93}. A benchmark verification was the
bounded retransmission protocol by
Philips~\cite{DBLP:conf/amast/GrooteP96, DBLP:conf/types/HelminkSV93},
which later turned out to be part of the Remote Control~6 standard for
infrared communication. The verification efforts unveiled serious
problems in extensions of the protocol and showed that deadlocks can
be overlooked when applying verification based on traces
\cite{DBLP:conf/types/HelminkSV93}.

Later we modelled and analysed far more systems: a distributed system
to lift large objects \cite{DBLP:journals/jlp/GrootePW03}, the data
link layer of the firewire protocol
\cite{DBLP:journals/corr/abs-2403-18723}, the Atacama large millimeter
array \cite{Plo09:tue}, a pacemaker \cite{Wig07:msc}, the Flexray
protocol \cite{Cra12:fmics}, heart-beat protocols \cite{Atif2011}, and
many more.  Our experience was both sobering and reassuring. For
virtually all systems, distributed algorithms and protocols that we
modelled and analysed, problems were uncovered.

From these specification and verification activities, we drew three
important observations. The first one was that formal specification
and verification boosts both quality and efficiency with a factor 10
and~3, respectively~\cite{DBLP:journals/sttt/OsaiweranSHGR16}. These
results have been obtained using the language ASD by
Verum~\cite{Bro05:fm}, a company that now uses mCRL2 as their
verification engine~\cite{DBLP:conf/fmics/BeusekomGHHWWW17}. The
reason for this remarkable result, comprising the second observation,
is that by specifying properties and checking these on the behavioural
model, different perspectives on the same system are confronted with
each other, leading to a multiplication of error probability causing
the overall error probability to become much
lower~\cite{DBLP:journals/scp/BrandG15}. The third observation is that
in order to keep the state space in check, the way models are
specified is very important~\cite{DBLP:journals/stvr/GrooteKO15}.
This is a somewhat neglected field in the realm of formal methods where
algorithms and formalisms receive more attention than their usage.
However, with the right modelling skills it is possible to model and
analyse controllers of truly immense systems \cite{BBGBB25:fmics,
  groote2025formalspecificationdesiredsoftware,
  DBLP:journals/corr/abs-2403-18722}, and these specifications can
play the role of their design documents or software blueprints.

We currently see a new trend in modelling and verification.  As it is
relatively easy to design a language, there are many formalisms being
developed to model in a certain application domain.  These formalisms
can all benefit from model checking, but it has become too expensive
to develop a dedicated model checker or similar verification tool.  It
is more efficient to build a translator to an existing model checker
and use that.  There are translators to mCRL2 for UML, SysML, Rebecca,
CIF, Reo, Dezyne, Cordis-UML \cite{DBLP:journals/isse/HansenKLMP10,
  DBLP:conf/forte/BouwmanLW21, DBLP:conf/acsd/HojjatSMG07,
  DBLP:conf/case/ReniersK24, KKV12:fac, JCP12:foclasa,
  DBLP:conf/fmics/BeusekomGHHWWW17,
  DBLP:conf/fmics/StramagliaK22}. Such translators have been used for
instance to systematically analyse railway
problems~\cite{DBLP:journals/fac/BouwmanWLSR23}.

\section{Future developments}

How will we proceed with mCRL2 from here? At the moment we see the
following points that should be addressed.
\begin{itemize}
\item Although it is possible to analyse large systems, it would be
  nice to increase the analysis capabilities of the tools. It is
  expected that this goes along the line of improving symbolic model
  checking, either by finding compression techniques better than BDDs,
  or employing BDDs better by identifying more precisely which
  parameters in linear processes and PBESs are independent.
\item As already indicated, very little attention has been paid to the
  ideal style of specification, to keep the state space in check. We
  expect that with larger applications, the relevance of this question
  will become more prominent.
\item With the specification of the software of the Maeslant barrier
  and road tunnels \cite{BBGBB25:fmics,
    DBLP:journals/corr/abs-2403-18722} for example, it sometimes feels
  that mCRL2 can be optimised as a language, allowing to make the
  models much more compact without losing comprehensibility. Should
  we adapt the language, with the risk of damaging its mathematical
  underpinning?  For a comparable development one can look at the
  transition from LOTOS to LNT~\cite{DBLP:conf/birthday/GaravelLS17}.
\item The language mCRL2 supports time and to a certain extent
  probabilities. But the semantics of especially continuous
  probability distributions and non-deterministic choice is not yet
  fully understood in the context of
  mCRL2~\cite{DBLP:journals/corr/abs-2108-10489}. Furthermore,
  analysis techniques for time and probabilities, e.g., methods to
  solve PRESs, can and should be improved. There is also a long
  standing desire~\cite{Cui04:phd} to merge discrete process theories
  with control theory and cyber-physical systems \cite{Alu15:mit,
    Pla18:springer}, but despite attempts little progress has been
  made in this direction.
\item In the area of applications, it would be interesting to analyse
  the reliability of software controlled systems using probabilistic
  process algebras~\cite{BG25:rams}. As it stands it is not really
  possible to quantitatively predict the reliability of the software
  in safety-critical systems in the same way as the low probabilities
  of failure of hardware infrastructure can be assessed.
\end{itemize}

\section{Conclusion}

Looking back, mCRL2 and its associated toolset can be considered an
impressive accomplishment.
Firstly, effective and expressive modelling formalisms have been
defined both to describe the behaviour for systems that communicate by
exchanging messages, and for denoting properties for such systems.
Secondly, a wide range of analysis methodologies, algorithms, and
tools have been made available that help to provide insight into the
systems under consideration. And thirdly, quite a number of
applications have been studied that show that this framework is really
useful.

Although the outcome of steady progress, we only could come this far
because, in retrospect, we always adhered to two guiding
principles. Namely, as first principle, whatever we do, it must be
mathematically well-founded and sound, and, as second principle, we
should be able to get insight in the behaviour of actual
systems. \added{This second principle often guided the way forward,
  while the first ensured that we did things correctly.}

There are, of course, also aspects that we did not achieve. There are
still many unanswered questions, both of a theoretical and practical
nature as discussed in the previous section. But maybe more
importantly, there is not yet a common formalism to model the
behaviour of systems. Compare this to the situation in control theory
where differential equations have become the lingua franca in
modelling, analysing, and designing physical systems. Scientists and
engineers all over the world share this as the standard framework.

Maybe it comes too soon to desire a widely established formalism for
the modeling of system behaviour for now, as we do not even know into
which directions computer science and computer programming develop.
The future is hard to predict, given that the capabilities of
digital hardware are still growing exponentially, and techniques such
as AI may drastically change our way of working.
Still, we believe that for many decades to come, the
construction of computer controlled systems will benefit from the
availability of formal modelling and analysis techniques.

\begin{acks}
  The authors are grateful to Jeroen Keiren, Bas Luttik, Tim Willemse
  and to the referees for their helpful comments on this paper and
  their assistance with completing the bibliography.
\end{acks}

\bibliographystyle{ACM-Reference-Format}
\bibliography{references}

\end{document}